\documentclass[Entropy,article,accept,moreauthors,pdftex,12pt,a4paper]{mdpi} 
\lastpage{x}
\doinum{10.3390/------}
\pubvolume{xx}
\pubyear{2013}
\history{Received: 1 April 2013; in revised form: 29 May 2013 / Accepted: 1 June 2013 / \\
Published: xx June 2013}

\usepackage{amssymb,amsmath}
\usepackage{graphicx}
\usepackage{subfigure}
\makeatletter
\renewcommand{\@thesubfigure}{\normalsize(\textbf{\alph{subfigure}})}
\makeatother
\usepackage[T1]{fontenc}
\usepackage{verbatim}
\usepackage{soul,booktabs}
\usepackage{cancel}
\usepackage[normalem]{ulem}
\usepackage{xcolor}
\usepackage{varwidth}
\usepackage{hyphenat}
\Title{Quantum Contextuality with Stabilizer States}

\Author{Mark Howard *, Eoin Brennan and Jiri Vala }

\address[1]{%
Department of Mathematical Physics, National University of Ireland, Maynooth, Ireland; \linebreak{E-Mails: eoin.brennan@nuim.ie (E.B.); jiri.vala@nuim.ie (J.V.)}}

\corres{E-Mail:~mark.howard@nuim.ie; \linebreak {Tel./Fax}. +353 1 708 3339/+353 1 708 3967}

\abstract{The Pauli groups are ubiquitous in quantum information theory because of their usefulness in describing quantum states and operations and their readily \linebreak understood symmetry properties. In addition, the most well-understood quantum error \linebreak correcting codes---stabilizer codes---are built using Pauli operators. The eigenstates of these \linebreak operators---stabilizer states---display a structure (e.g., mutual orthogonality relationships) that has made them useful in examples of multi-qubit non-locality and contextuality. Here, we apply the graph-theoretical contextuality formalism of Cabello, Severini and Winter to sets of stabilizer states, with particular attention to the effect of generalizing two-level qubit systems to {odd prime}
$d$-level qudit systems. 
{While} state-independent contextuality using two-qubit states does not generalize to qudits, 
we show explicitly how state-dependent contextuality associated with a Bell inequality does generalize. Along the way we note various structural properties of stabilizer states, with respect to their orthogonality relationships, which may be of independent interest. }

\keyword{contextuality; stabilizer; non-locality}

\DeclareMathOperator{\Tr}{Tr}
\def\ket #1{\vert #1\rangle}
\def\bra #1{\langle #1\vert}
\newcommand{\ketbra}[2]{\ensuremath{\ket{#1}\!\bra{#2}}}
\newcommand{\Jam}{Jamio\l kowski }

\newcommand{\Zd}{\ensuremath{\mathbb{Z}_d}}

\newcommand{\SL}{\mathrm{SL}}

\begin{document}

\section{Introduction}

The Pauli operators are ubiquitous in quantum information theory, typically used as an operator basis to decompose multi-particle states or circuits. There are many combinatorial and geometrical structures that arise in finite-dimensional quantum mechanics that are intimately related with 
Pauli operators, and these structures are known to have favourable properties for use in quantum information processing. Examples include Mutually Unbiased Bases (MUBs \cite{Bandyopadhyay2002,Lawrence:2002}), Weyl--Heisenberg covariant Symmetric Informationally-Complete Positive Operator Value Measures (SIC-POVMs \cite{Rennes:2004}) and spherical/unitary designs \cite{Scott:2008,WvDMH:2011}. While the qubit (two-level system) versions of the
{Pauli} operators are the most familiar, the generalization to qudit ($d$-level systems) Pauli operators is {mathematically} straightforward. {What is not straightforward, however, is whether construction techniques, like those in the works listed above, will generalize and retain their desirable properties when we switch to general qudit systems.}

With regard to fault-tolerant universal quantum computation (UQC), Pauli operators arise naturally within the context of stabilizer error-correcting codes and so they are intimately related to this task too. On the other hand, the Gottesman--Knill theorem tells us we cannot see better-than-classical computational performance using quantum circuits restricted to (i) operating on Pauli eigenstates, (ii) using Pauli measurements; and (iii) using gates that inter-convert Pauli operators (Clifford gates). {In this model of fault-tolerant computation, the use of qudits appears to offer an advantage over qubits in terms of the efficiency associated with a magic state distillation routine \cite{Campbell:2012}.}

In this work we examine whether sets of stabilizer states (\emph{i.e}., Pauli eigenstates) exhibit quantum contextuality. More precisely, we examine whether known instances of quantum contextuality for \linebreak two-qubit states generalize to two-qudit states {when $d$ is an odd prime}. The most directly comparable work is by Planat \cite{Planat:arxiv12} although he restricts his search for contextuality to the subset of stabilizer states that can be represented as real vectors and allows non-power-of-prime dimensions.

Because $d=2$ is the only even prime dimension, we sometimes see striking dissimilarities in combinatorial or geometrical structures, depending on whether the prime Hilbert space dimension is $d=2$ or $d>2$. An example of this difference can be seen in the context of a discrete Wigner function (DWF) where the associated geometrical structure is simplified in (power-of-) odd prime dimensions \cite{Gross:2006,Veitch:2012}. We will see a similar distinction when we seek to find instances of state-independent contextuality; such instances are fairly easy to concoct using multiple qubits but seem 
impossible 
to find using multiple {odd-prime dimensional} qudits.

Nonlocality can be understood as a special type of contextuality, and so Bell inequalities can be recast in the graph-based contextuality formalism. We derive such a decomposition for a family of two-qudit Bell inequalities and discuss the related orthogonality graphs, graph parameters and state-dependent noncontextuality inequalities.

We begin by providing the necessary definitions and mathematical background along with a discussion of quantum contextuality. The subsequent results section is partitioned into three main sections. The first two subsections relate to state-independent and state-dependent contextuality respectively. The remaining subsection provides some additional considerations related to graph-based contextuality that should be borne in mind, and discusses how these apply to our current investigation.

\section{Mathematical Preliminaries}\label{sec:Mathematical Preliminaries}

In order to appreciate the results of Section~\ref{sec:Results} we will first need to provide the necessary mathematical background and definitions. We begin by discussing the generalized Pauli group, the associated stabilizer states and the unitary operations that inter-convert them. In the next subsection we introduce quantum contextuality and its graph-theoretical interpretation. This latter aspect of our work requires a dedicated subsection that lists all the purely graph-theoretical relationships and results that are used elsewhere in the paper.

\subsection{Stabilizer States and Clifford Gates}\label{sec:Stabilizer States and Clifford Gates}
\vspace{-12pt}

\subsubsection{. Stabilizer States}\label{sec:Stabilizer States}
Throughout, we always assume the dimension, $d$, of a single qudit to be a prime number. Because of the mathematics involved, most of the subscripts, superscripts and arithmetic will use elements \linebreak from $\Zd$---the set of integers modulo $d$. Occasionally we will also need to use $\Zd^*$---the set of non-zero integers modulo $d$. Define the generalized Pauli ``shift'' and ``phase'' operators, respectively, as
\begin{align}
X\ket{j}=\ket{j+1}, \quad Z\ket{j}=\omega^j\ket{j} \quad \text{where} \quad\omega= \exp(2 \pi i/d)
\end{align}
so that in the $d=2$ case these reduce to the familiar $\sigma_x$ and $\sigma_z$ operators. Taking all products and powers of these $X$ and $Z$ operators produces the set of Pauli operators
\begin{align}
P_{(x|z)}=\begin{cases} i^{xz}X^xZ^z \quad (d=2) \\ \quad X^xZ^z \quad(d>2) \end{cases}
\end{align}
where we have introduced a phase in the $d=2$ case that ensures $P_{(1|1)}$ coincides with our normal definition of the Pauli $Y=\sigma_y$ operator. The Pauli group $\mathcal{G}$ for a single particle is comprised of the Pauli operators along with a set of global phases,
\begin{align}
\mathcal{G}=\begin{cases} \left\{ i^k P_{(x|z)} \mid k\in \mathbb{Z}_4, x,z \in \mathbb{Z}_2 \right\} &\quad (d=2) \\ \left\{ \omega^k P_{(x|z)} \mid k,x,z \in \mathbb{Z}_d \right\} &\quad(d>2) \end{cases} \label{eqn:PauliGroup}
\end{align}

Using so-called symplectic notation, the general form for an $n$-particle Pauli operator is
\begin{align}
P_{(x|z)}=\left(X^{x_1}\otimes X^{x_2}\ldots\right) \left(Z^{z_1}\otimes Z^{z_2}\ldots\right)\
\end{align}
where $x$ and $z$ are now vectors of length $n$ \emph{i.e}., $x,z \in \Zd^n$, where each $x_i$ or $z_j$ is an element of $\Zd$. Two operators $P_{(x|z)}$ and $P_{(x^\prime|z^\prime)}$ commute if and only if the symplectic inner product between their vector descriptions vanishes \emph{i.e}., if and only if $\protect{\sum_i (x_iz^\prime_i-x^\prime_iz_i)=0}$.

%


Because of the graph-based contextuality formalism, we are particularly interested in the projectors associated with the Pauli operators. The $\omega^k$ eigenspace of a single-qudit operator $P_{(x_1|z_1)}$ corresponds to the projector
\begin{align}
\Pi_{(x_1|z_1)[k]}=\sum_{j\in \Zd}\omega^{-jk}(P_{(x_1|z_1)})^j \label{eqn:Qproj}
\end{align}
which is clearly a qudit stabilizer state \emph{i.e}., it is a rank-$1$ stabilizer projector. There are a total of $d(d+1)$ distinct single-qudit stabilizer states \cite{Ivonovic:1981} in a Hilbert space of dimension $d$, and they can be identified as the eigenstates of the following set of operators
\begin{align}
\text{MUB operators:} \qquad \{P_{(0|1)},P_{(1|0)},P_{(1|1)},\ldots,P_{(1|d-1)}\} \label{eqn:MUBset}
\end{align}

Measuring a two-qudit Pauli operator corresponds to projecting with the following rank-$d$ operator
\begin{align}
\Pi_{(x_1,x_2|z_1,z_2)[k]}=\sum_{j\in \Zd}\omega^{-jk}(P_{(x_1,x_2|z_1,z_2)})^j \label{eqn:TwoQProj}
\end{align}
This projector can be decomposed into a sum of $d$ rank-1 projectors via,
\begin{align}
\Pi_{(x_1,x_2|z_1,z_2)[k]}=\sum_{a+b=k}\left(\Pi_{(x_1|z_1)[a]} \otimes \Pi_{(x_2|z_2)[b]}\right) \label{eqn:SumRank1}
\end{align}
where there are $d$ solutions $(a,b)$ for a given $k$ in the summation. To prove this last identity, one must insert the definitions given in Equation~\eqref{eqn:Qproj} and Equation~\eqref{eqn:TwoQProj} and use the fact that
\begin{align}
\sum_{\substack{m,n, \\ a+b=k}}\omega^{-(ma+nb)}A^m \otimes B^n=d\sum_{n}\omega^{-nk}(A\otimes B)^n \quad \text{since } \sum_{a\in \Zd}\omega^{a(m-n)}=d\delta_{m,n}
\end{align}

The number of distinct two-qudit stabilizer states (see e.g., \cite{Gross:2006}) is
\begin{align}
\{\Pi\}_{\textrm{tot}}=d^2(d^2+1)(d+1) \qquad \text{(e.g., }d=2:60, d=3:360, d=5:3900\text{)}
\end{align}
Of these states, the number of separable two-qudit states is
\begin{align}
\{\Pi\}_{\textrm{sep}}=[d(d+1)]^2\qquad \text{(e.g., }d=2:36, d=3:144, d=5:900\text{)}
\end{align}
which implies that the number of entangled two-qudit states is
\begin{align}
\{\Pi\}_{\textrm{ent}}=d^3(d^2-1)\qquad \text{(e.g., }d=2:24, d=3:216, d=5:3000\text{)}
\end{align}
Later, we will be interested in whether these sets of stabilizer states exhibit state-independent contextuality.

\subsubsection{. The Clifford Group}

The set of unitary operators that map the qudit Pauli group onto itself under conjugation is called the Clifford group, $\mathcal{C}_d$,
\begin{align*}
\mathcal{C}_d=\{C\in U(d)\vert C P_{(x|z)} C^\dag \propto P_{(x|z)}^\prime\}
\end{align*}
where the proportionality symbol denotes equality up to a phase of the form $\omega^k$ (or $i^k$ in the case of qubits). The number of distinct Clifford gates for a single qudit system is $|\mathcal{C}_d|=d^3(d^2-1)$, and this can be seen by noting the isomorphism
\begin{align}
\mathcal{C}_d \cong \SL(2,\Zd) \ltimes \Zd^2
\end{align}
established by Appleby \cite{Appleby:arxiv09}. If we specify the elements of $\SL(2,\Zd) $ and $\Zd^2$ as
\begin{align}
F=\left(
\begin{array}{cc}
\alpha & \beta \\
\gamma & \delta \\
\end{array}
\right)\in \SL(2,\Zd) \qquad u=\left(
\begin{array}{c}
u_1 \\
u_2 \\
\end{array}
\right)\in \Zd^2 \label{Fudef}
\end{align}
then \cite{Appleby:arxiv09} provides an explicit description of the unitary matrix $\protect{C_{(F\vert u)} \in \mathcal{C}_d}$ in terms of these elements~\emph{i.e}.,
\begin{align}
&C_{(F\vert u)}=P_{(u_1|u_2)}U_F \\
&U_F=\begin{cases}
\frac{1}{\sqrt{p}}\sum_{j,k=0}^{p-1}\tau^{\beta^{-1}\left(\alpha k^2-2 j k +\delta j^2\right)}\ket{j}\bra{k}\quad &\beta\neq0\\
\sum_{k=0}^{p-1} \tau^{\alpha \gamma k^2} \ket{\alpha k}\bra{k}\quad &\beta=0
\end{cases}
\end{align}
where $\tau=\omega^{2^{-1}}$ if $d>2$ and $\tau=(-i)$ if $d=2$. 

The canonical two-qudit Bell state,
\begin{align}
\ket{\Phi}=\frac{1}{\sqrt{d}}\sum_{j=0}^{j=d-1} \ket{jj}
\end{align}
is the unique $+1$ eigenstate of $d^2$ mutually commuting Pauli operators, powers and products of $X\otimes X$ and $Z\otimes Z^{-1}$ \emph{i.e}.,
\begin{align}
\ketbra{\Phi}{\Phi}&=\frac{1}{d^2} \left(I\otimes I+X\otimes X+\ldots+(X\otimes X)^{d-1}\right)\left(I\otimes I+Z\otimes Z^{-1}+\ldots+(Z\otimes Z^{-1})^{d-1} \right)
\end{align}
A \Jam state, $\ket{J_U} \in \mathbb{C}^{d^2}$, corresponding to a unitary operation $U \in \mathrm{U}(d)$ is defined by
\begin{align}
&\ket{J_U}=(\mathbb{I}\otimes U)\ket{\Phi} \label{eqn:JamIso}
\end{align}
This correspondence between operations and higher-dimensional states is known as the \Jam isomorphism. If the unitary, $U$, used in Equation~\eqref{eqn:JamIso} is a Clifford operation, then the resulting \Jam state is a bipartite entangled stabilizer (BES) state. In fact, any BES state must be a \Jam isomorph of a Clifford gate so that the number of BES states is
\begin{align}
|\{\Pi\}_{\textrm{ent}}|=|\mathcal{C}_d|=d^3(d^2-1)
\end{align}
We will use this group structure associated with BES states later, when we discuss whether the set $|\{\Pi\}_{\textrm{ent}}|$ exhibits contextuality.

\subsection{Quantum Contextuality and Graph Theory}\label{sec:Quantum Contextuality and Graph Theory}

The following argument, as put forward by Mermin \cite{Mermin:1990}, helps illustrate the counterintuitive nature of quantum contextuality. Consider a set of mutually commuting operators $\{A,B,C,\ldots\}$ and attempt to ascribe to each measurement a value $\{\nu(A)\in \lambda(A),\nu(B)\in \lambda(B),\nu(C)\in \lambda(C),\ldots\}$ respectively, where $\lambda(A)$ denotes the spectrum of $A$ and so on. If some functional relation $f$ exists, such that
\begin{align}
f(A,B,C,\ldots)=0
\end{align}
is an operator identity, then we should also have that
\begin{align}
f(\nu(A),\nu(B),\nu(C),\ldots)=0
\end{align}
Explicitly providing sets of operators for which this is not true shows that our assumption of assigning $\nu(A)$ as the outcome of measurement $A$ \emph{etc}.~was unjustified. In other words, we cannot assume that quantum measurement is a process that just reveals pre-existing properties of a state, independently of any other compatible measurements that are being performed.

Arguably, the canonical example of quantum contextuality is provided by the Peres--Mermin magic square \cite{Mermin:1990,Peres:1991}. Here, the mutually commuting sets of operators are triples of two-qubit Pauli operators. There are $6$ sets of commuting operators, where each set corresponds to a row or column of Table \ref{tab:PMsquare}. Define the operator comprised of the product of all three operators in Row $i$ or Column $j$ as $R_i$ and $C_j$ respectively. Each individual two-qubit Pauli operator has eigenvalue $\pm 1$, whereas it can be verified that $R_i=\mathbb{I}_4=-C_j$ for all $ i,j \in \{1,2,3\}$. This forces us to accept that $\nu(R_i)=1=-\nu(C_j)$. The only way $\nu(R_i)=1$ could hold is to have an even number of $-1$ assignments in each row. The only way $\nu(C_j)=-1$ could hold is to have an odd number of $-1$ assignments in each column. These last two statements are mutually contradictory---no such assignment can be found.

\begin{table}[H]
\centering
\begin{tabular}{cccc}
\toprule
\multicolumn{1}{c}{} & \multicolumn{1}{c}{\boldmath$C_1$} & \multicolumn{1}{c}{\boldmath$C_2$} & \multicolumn{1}{c}{\boldmath$C_3$} \\
\midrule $R_1$ & $X\otimes Y$ & $Y\otimes X$ & $Z\otimes Z$ \\
$R_2$ & $Y\otimes Z$ & $Z\otimes Y$ & $X\otimes X$ \\
$R_3$ & $Z\otimes X$ & $X\otimes Z$ & $Y\otimes Y$ \\
\bottomrule
\end{tabular}
\caption{\label{tab:PMsquare} One example of a Peres--Mermin (PM) magic square construction provided by Aravind \cite{Aravind:2006}. The operators in each cell of this table are two-qubit Pauli operators. If we try to assign $\pm 1$ values (\emph{i.e}., measurement outcomes) to each operator in a consistent way, we are forced into a contradiction.
}
\end{table}

Cabello, Severini and Winter \cite{CSW:arxiv2010} have recently introduced a graph-theoretical generalization of these Kochen--Specker type constructions, inspired by an earlier result of Klyachko \emph{et al.} \cite{KCBS:2008}. Consider a set of binary yes-no tests, which we quantum mechanically represent by a set of rank-one projectors, $\Pi$, with eigenvalues $\lambda(\Pi)\in\{1,0\}$. Compatible tests are those whose representative projectors commute, and a context is a set of mutually compatible tests. Contradictions can be found if we try to assign outcomes to these tests independently of their context.

Non-locality is a special instance of contextuality wherein compatibility of tests is enforced by spatial separation. The two-qubit CHSH inequality provides an example of a non-contextuality inequality, which in this case rules out a local non-contextual hidden variable theory. If we drop the requirement of spatial separation, then a single qutrit suffices to rule out non-contextual hidden variable theories, as we now discuss.

Take a set of five yes-no questions $\{\Pi_0,\Pi_1\ldots,\Pi_4\}$ such that $[\Pi_i,\Pi_{i\oplus 1}]=0$ where addition is performed modulo 5. We further impose, as usual, that commuting rank-1 projectors cannot both take on the value $+1$ \emph{i.e}., the respective propositions are mutually exclusive and cannot both be answered in the affirmative.
Graphically, we can represent this scenario as a pentagonal graph $\Gamma$ (\emph{i.e}., a $5$-cycle) where vertices correspond to tests, and adjacent (connected) vertices correspond to compatible and exclusive tests. Define the operator $\Sigma_{\Gamma}$ to be
\begin{align}
\Sigma_{\Gamma}=\sum_{\Pi \in \Gamma} \Pi
\end{align}
so that in the classical (non-contextual) case $\Sigma_{\Gamma}$ counts the number of yes (\emph{i.e}., $+1$) answers to the tests. The maximum number consistent with the rules outlined above is
\begin{align}
\langle \Sigma_{\Gamma} \rangle^{\textsc{NCHV}}_{\max}=2 \label{eqn:Sigmamax1}
\end{align}

\begin{figure}[H]
\centering
\includegraphics[scale=1.1]{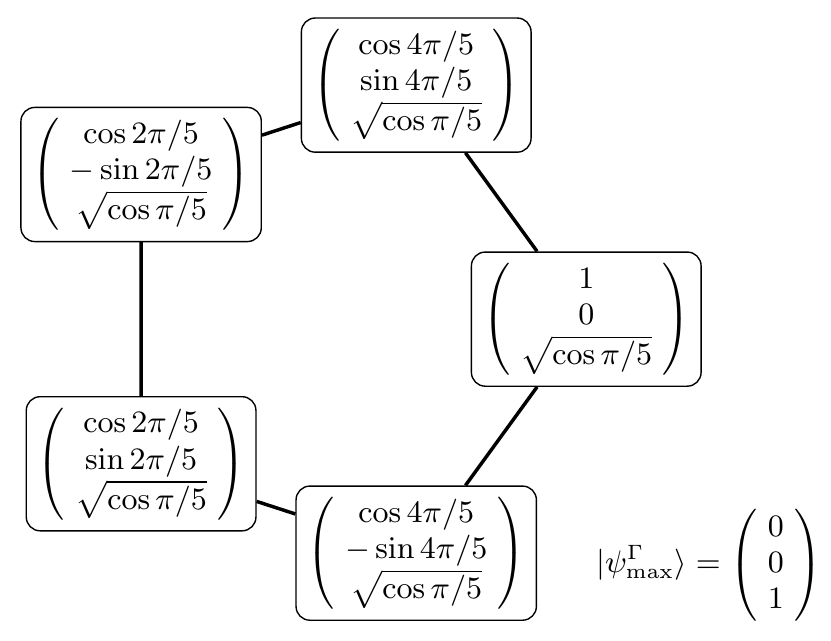}
\caption{\label{fig:5cycle} The KCBS contextuality construction \cite{KCBS:2008} involves five projectors of the form $\Pi_i = \ketbra{\psi_i}{\psi_i} \in \mathcal{H}_3$, where the un-normalized versions of $\ket{\psi_i}$ and their mutual orthogonality relations are given in the graph depicted (connected vertices correspond to compatible and exclusive tests). The state $\ket{\psi_{\max}^\Gamma}$ is maximally contextual, insofar as it maximally violates the noncontextuality inequality $\langle \Sigma_{\Gamma} \rangle^{\textsc{NCHV}}_{\max} \leq 2$.
}
\end{figure}

Quantum mechanically we can achieve an orthogonal representation of this graph using five vectors $\ket{\psi_{i}} \in \mathbb{C}^3$ (see Figure~\ref{fig:5cycle}) so that $[\Pi_i,\Pi_j]=0 \iff \langle \psi_i \ket{\psi_j}=0$ \emph{i.e}., $\Pi_i=\ketbra{\psi_i }{\psi_i }$. It was shown in~\cite{KCBS:2008,CSW:arxiv2010} that the maximum value of $\Sigma_{\Gamma}$ achievable by this qutrit representation is equal to the maximal achievable by arbitrary-dimensional quantum systems, \emph{i.e}.,
\begin{align}
\langle \Sigma_{\Gamma} \rangle^{\textsc{QM}}_{\max}=\sqrt{5}\approx 2.236 \label{eqn:Sigmamax2}
\end{align}
To see this, take the maximizing eigenvector $\ket{\psi_{\max}^\Gamma}$ of the Hermitian operator $\Sigma_{\Gamma}$ and note that
\begin{align}
\Tr(\Sigma_{\Gamma} \ketbra{\psi_{\max}^\Gamma}{\psi_{\max}^\Gamma})=\langle \Sigma_{\Gamma} \rangle^{\textsc{QM}}_{\max}=\sqrt{5}
\end{align}
The operator $\Sigma_{\Gamma}$ can be interpreted as a witness for contextuality, in the same way that Bell inequalities are witnesses for nonlocality \emph{i.e}.,
\begin{align}
\Tr(\Sigma_{\Gamma} \rho)>2 \Rightarrow \rho \text{ exhibits contextuality}
\end{align}
in the preceding 5-cycle example.

In the same way that nonlocality can be generalized to a general non-signalling theory (the most famous example being the non-local (Popescu--Rohrlich \cite{PR:1994}) box construction), contextuality can be broadened to include generalized probabilistic theories (GPT). Once again, the quantity $\Sigma_{\Gamma}$ can be maximized in a meaningful way e.g., in the preceding 5-cycle example we have
\begin{align}
\langle \Sigma_{\Gamma} \rangle^{\textsc{GPT}}_{\max}=\frac{5}{2} \label{eqn:Sigmamax3}
\end{align}

The remarkable result of \cite{CSW:arxiv2010} is that the quantities $\langle \Sigma_{\Gamma} \rangle^{\textsc{X}}_{\max}$ for $\textsc{X} \in \{\textsc{NCHV,QM,GPT}\}$ and for a general {orthogonality} graph correspond to well-known graph-theoretical quantities \emph{i.e}., $\{\alpha(\Gamma),\vartheta(\Gamma),\alpha^*(\Gamma)\}$ respectively. In the 5-cycle example these quantities equate to values given in Equations \eqref{eqn:Sigmamax1}, \eqref{eqn:Sigmamax2} and \eqref{eqn:Sigmamax3}. We can use the many relationships between these and other graph-theoretical quantities e.g.,
\begin{align}
\alpha(\Gamma)\leq \vartheta(\Gamma)\leq \alpha^*(\Gamma)
\end{align}
which is true for all graphs, and this implies
\begin{align}
\langle \Sigma_{\Gamma} \rangle^{\textsc{NCHV}}_{\max}\leq\langle \Sigma_{\Gamma} \rangle^{\textsc{QM}}_{\max} \leq \langle \Sigma_{\Gamma} \rangle^{\textsc{GPT}}_{\max}
\end{align}
which makes intuitive sense. Graphs for which $\protect{\alpha(\Gamma) < \vartheta(\Gamma)}$ indicate that appropriately chosen projectors $\Pi$ can reveal quantum contextuality. It is important to note that not all realizations of $\Gamma$ using a set of projectors will reach the maximum value $\vartheta(\Gamma)$. Instead, $\vartheta(\Gamma)$ is an upper bound that can always be reached, by optimizing over the set of projectors used and by optimizing the state $\rho$ whose expectation value is $\langle \Sigma_{\Gamma} \rangle^{\textsc{QM}}=\Tr(\rho\Sigma_{\Gamma}) $ with respect to the operator $\Sigma_{\Gamma}$. Sadiq \emph{et al.} \cite{Sadiq:2013} discuss this non-optimality in the context of nonlocality, where they argue that spacelike separation is not fully captured by the orthogonality graphs as we have described. We will see similar manifestations of this non-optimality in Section~\ref{sec:State-dependent contextuality using Bell inequalities}
where the expectation value $\langle \Sigma_{\Gamma} \rangle$ is significantly less than the most general upper bound given by $\vartheta(\Gamma)$.

%

One significant difference between nonlocality and general contextuality is that arbitrary (even maximally mixed) states can exhibit contextuality when the set of tests comprise an example of state-independent contextuality (SIC), e.g., the projectors associated with the Peres--Mermin magic square \cite{Mermin:1990,Peres:1991}. A necessary but not sufficient condition \cite{Cabello:arxiv2011} for a collection of tests $\{\Pi\}$ to exhibit SIC is that $\chi(\Gamma)>D$, where $\chi(\Gamma)$ is the chromatic number (explained in Section~\ref{sec:Graphtheoreticalprerequisites}) of the orthogonality graph $\Gamma$ and $D$ is dimensionality of the Hilbert space projectors \emph{i.e}., $\Pi \in \mathcal{H}_D$.


\subsubsection{. Graph Theoretical Prerequisites \label{sec:Graphtheoreticalprerequisites}}
In order to prove most of our claims we will need to appeal to some well-known facts and results from graph theory. Here we provide the necessary background. A graph, $\Gamma$, is a mathematical structure consisting of a set of vertices, and a set of edges connecting the vertices. 
If two vertices $g$ and $h$ are adjacent (connected) then we denote that as $g \sim h$. Since we only ever consider undirected graphs, this is equivalent to $h\sim g$. The graph complement $\overline{\Gamma}$ of a graph $\Gamma$ is obtained by replacing edges with non-edges and vice-versa. An independent set of a graph is a set of vertices, no two of which are adjacent. The independence number $\alpha(\Gamma) \in \mathbb{N}$ is the size of the largest such set. The clique number $\omega(\Gamma) \in \mathbb{N}$ is the size of largest set of vertices in which every member is connected to every other member. The clique number of any graph is equal to the independence number of its complement and vice versa \emph{i.e}., $\omega(\Gamma)=\alpha(\overline{\Gamma})$. The Lovasz theta number $\vartheta(\Gamma) \in \mathbb{R}$ is the solution of a certain semi-definite programming problem while the fractional packing number $\alpha^*(\Gamma) \in \mathbb{Q}$ is the solution of a certain linear program (see~\cite{CSW:arxiv2010} for details). The integer $\bar{\chi}(\Gamma) \in \mathbb{N}$---the so-called clique cover number---represents the minimum number of cliques needed to cover every vertex of $\Gamma$. It is well-known that this serves as an upper bound for the fractional packing number \emph{i.e}., $\alpha^*(\Gamma) \leq \bar{\chi}(\Gamma)$. The vertex coloring problem for a graph involves assigning a color to every vertex in such a way that adjacent vertices cannot be assigned the same color. The minimum number of colors required to do this is the chromatic number $\chi(\Gamma)$. A little thought reveals that the chromatic number is bound below by the clique number \emph{i.e}., $\omega(\Gamma) \leq \chi(\Gamma)$.

An undirected Cayley graph $\Gamma(G,T)$ with an associated finite group $G$ and set $T \subset G$ is the graph whose vertices are the elements of $G$ and whose set of edges is $\{g\sim h|g^{-1}h \in T\}$. For an undirected Cayley graph without self-loops we must have $I \not\in T$ and $T^{-1}=T$. The resulting graph $\Gamma(G,T)$ is regular \emph{i.e}., each vertex is adjacent to $|T|$ other vertices. 

Given two graphs $G$ and $H$ with respective vertex sets $\{g,g^\prime, \ldots \}$ and $\{h,h^\prime, \ldots \}$, we can define various graph products whose vertex set is the Cartesian product of $\{g,g^\prime, \ldots \}$ and $\{h,h^\prime, \ldots \}$ but where the condition $(g,h) \sim (g^\prime,h^\prime)$ can be defined in a number of different ways. For example the \textsc{OR} product was already used in \cite{Cabello:2013} in the context of quantum contextuality
\begin{align}
G \textsc{ OR } H: \quad (g,h) \sim (g^\prime,h^\prime) \iff g \sim g^\prime \text{ or } h \sim h^\prime \label{eqn:ORdefn}
\end{align}


There are also distinct notions of graph addition, the simplest being the disjoint union whereby the total vertex set is the union of vertex sets and the total edge set is the union of edge sets. We denote the disjoint union of $m$ copies of the complete graph $K_n$ as $mK_n$

\section{Results}\label{sec:Results}
\vspace{-12pt}
\subsection{State-Independent Contextuality Using Stabilizer States}\label{sec:State-independent contextuality using stabilizer states}

In this section, we will {apply graph-theoretical techniques} to sets of stabilizer states to see whether they are sufficient to exhibit state-independent contextuality (SIC). We begin by concentrating on the set of bipartite separable stabilizer states, $\{\Pi\}_{\textrm{sep}}$, and show that they are insufficient for exhibiting SIC. We next concentrate on the subset of two-qudit stabilizer states that are maximally entangled, $\{\Pi\}_{\textrm{ent}}$. As we discuss below, this subset is isomorphic to the Clifford group, hence powerful group-theoretical techniques can be used to examine the associated {orthogonality} graph $\Gamma_{\textrm{ent}}$. Finally, we examine the total (separable plus entangled) set of bipartite stabilizer states, $\{\Pi\}_{\textrm{tot}}$, and show that it cannot exhibit SIC for qudits of small odd prime dimension. Tables \ref{tab:GraphOrders}--\ref{tab:GraphExpectations} summarize our results concerning the structure of sets of projectors and whether they display contextuality. It seems clear that state-independent contextuality is much easier to achieve using qubits as opposed to qudits with {odd} prime $d>2$. {At a superficial level, one could} attribute this phenomenon to structural differences in the Pauli group as outlined in Equation~\eqref{eqn:PauliGroup}. Because the multi-qubit Pauli group necessitates a $d^2$-th \emph{i.e}., $4$-th root of unity in its phases, the structure is somewhat richer and non-contextual contradictions are easy to find. The simpler structure of the multi-qudit Pauli group for {prime} $d>2$ appears to make state-independent contextuality more difficult, if not impossible, to find. {We shall see in Section~\ref{sec:Combined separable and entangled stabilizer states} that the local hidden variable model provided by an appropriate discretized Wigner function provides a deeper and more general reason that forbids state-independent contextuality using stabilizer projectors in certain dimensions.}

\begin{table}[H]
\centering
\begin{tabular}{@{}cccc@{}}
\toprule
& \boldmath$|\Gamma_{\textrm{sep}}|$ & \boldmath$|\Gamma_{\textrm{ent}}|$ & \boldmath$|\Gamma_{\textrm{tot}}|$ \\
\midrule $d=2$ & 36 & 24 & 60 \\
$d=3$ & 144 & 216 & 360 \\
$d=5$ & 900 & 3000 & 3900 \\
$d$ & $[d(d+1)]^2$ & $d^3(d^2-1)$ & $d^2(d^2+1)(d+1)$ \\
\bottomrule
\end{tabular}
\caption{\label{tab:GraphOrders} The order (number of vertices) of the {orthogonality} graphs associated with \mbox{two-qudit} stabilizer states.
}
\end{table}

\begin{table}[H]
\centering
\begin{tabular}{@{}cccc@{}}
\toprule
& \boldmath$\alpha(\Gamma_{\textrm{sep}})$ & \boldmath$\alpha(\Gamma_{\textrm{ent}})$ & \boldmath$\alpha(\Gamma_{\textrm{tot}})$ \\
\midrule $d=2$ & $9$ & $\mathbf{5}$ & $\mathbf{12}$ \\
$d=3$ & $16$ & $24$ & $340$ \\
$d=5$ & $36$ & $120$ & $156$ \\
$d>2$ & $(d+1)^2$ & $d(d^2-1)^*$ & $(d^2+1)(d+1) ^*$ \\
\bottomrule
\end{tabular}
\caption{\label{tab:GraphAlphas} The independence number $\alpha$ of the {orthogonality} graphs associated with separable, entangled and all two-qudit stabilizer states. The quantities with {asterisks} are those for which we do not have a {graph-theoretical} proof, but which {the more general results referenced in Section~\ref{sec:Combined separable and entangled stabilizer states} support}. The contextuality formalism of \cite{CSW:arxiv2010} tells us that maximum value of $\langle \Sigma_\Gamma \rangle$ in a non-contextual hidden variable theory is given by $\alpha(\Gamma)$. The values in bold font are those that differ with the corresponding quantum expectation value given in Table \ref{tab:GraphExpectations}, and therefore the associated projectors exhibit contextuality.
}
\end{table}

\begin{table}[H]
\centering
\begin{tabular}{@{}cccc@{}}
\toprule
& \boldmath$\langle \Sigma_{\Gamma_{\textrm{sep}}} \rangle^{\textsc{QM}}=\overline{\chi}(\Gamma_{\textrm{sep}})$ & \boldmath$\langle \Sigma_{\Gamma_{\textrm{ent}}} \rangle^{\textsc{QM}}=\overline{\chi}(\Gamma_{\textrm{ent}})$ & \boldmath$\langle \Sigma_{\Gamma_{\textrm{tot}}} \rangle^{\textsc{QM}}=\overline{\chi}(\Gamma_{\textrm{tot}})$ \\
\midrule $d=2$ & 9 & 6 & 15 \\
$d=3$ & 16 & 24 & 340 \\
$d=5$ & 36 & 120 & 156 \\
$d$ & $(d+1)^2$ & $d(d^2-1)$ & $(d^2+1)(d+1)$ \\
\bottomrule
\end{tabular}
\caption{\label{tab:GraphExpectations} The achievable values of $\langle \Sigma_\Gamma \rangle$ in quantum mechanics, for each of the relevant sets of stabilizer projectors. This happens to coincide with the clique cover number $\overline{\chi}(\Gamma)$ of the related {orthogonality} graph. Comparing with the previous Table \ref{tab:GraphAlphas}, we see that the only situations for which quantum contextuality appears, because $\alpha(\Gamma) < \langle \Sigma_\Gamma \rangle^{\textsc{QM}}$, is when we include the set of entangled two-qubit states.
}
\end{table}

\subsubsection{. Separable Stabilizer States}

Associated with each of the $d+1$ MUB operators in Equation~\eqref{eqn:MUBset} is a complete orthonormal basis of $d$ states. Furthermore, states from different bases always have the same overlap. To identify individual states and bases we introduce the following notation
\begin{align}
\left\{\mathcal{B}^{\textrm{bas}}_{\textrm{vec}}\right\}= \left\{\mathcal{B}^1_1,\mathcal{B}^1_2,\mathcal{B}^1_3,\ldots,,\mathcal{B}^1_{d},\mathcal{B}^2_1,\ldots,\mathcal{B}^{d+1}_{d}\right\}
\end{align}
and the overlap between any two states is given by
\begin{align}
\Tr(\mathcal{B}^a_b \mathcal{B}^{a^\prime}_{b^\prime})=\frac{1}{d}(1-\delta_{a,a^\prime})+\delta_{a,a^\prime}\delta_{b,b^\prime}
\end{align}

Two stabilizer states $\mathcal{B}^a_b$ and $\mathcal{B}^{a^\prime}_{b^\prime}$ are orthogonal if and only if they are distinct members of the same~basis
\begin{align}
\Tr(\mathcal{B}^a_b \mathcal{B}^{a^\prime}_{b^\prime})=0 \iff a=a^\prime \text{ and } b\neq b^\prime
\end{align}

In order to examine the orthogonality graph $\Gamma_{\textrm{single}}$ of these single-qudit stabilizer states, we associate with each state $\mathcal{B}^a_b$ a vertex $v(a,b)$. Clearly there are $|\Gamma_{\textrm{single}}|=d(d+1)$ vertices with edges
\begin{align}
v(a,b)\sim v(a^\prime,b^\prime) \iff a=a^\prime \text{ and } b\neq b^\prime
\end{align}
so that $\Gamma_{\textrm{single}}$ is isomorphic to $d+1$ copies of the complete graph on $d$ vertices,
\begin{align}
\Gamma_{\textrm{single}}=(d+1)K_d
\end{align}

Maximum independent sets and maximum cliques are easy to write in this notation;
\begin{align}
\text{Maximum independent set:} \qquad \text{e.g., }\quad &\{v(a,1)|\ a \in \{1,2,\ldots,d+1\}\}\\
\text{Maximum clique:} \qquad \text{e.g., }\quad &\{v(1,b)|\ b \in \{1,2,\ldots,d\}\}
\end{align}

Next consider what happens if we take tensor products of single-qudit stabilizer states with each other. There are $[d(d+1)]^2$ separable two-qudit stabilizer states of the form $\mathcal{B}^{a_1}_{b_1} \otimes \mathcal{B}^{a_2}_{b_2}$, indexed by $a_1,a_2 \in \{1,2,\ldots, d+1\}$ and $b_1,b_2 \in \{1,2,\ldots, d\}$. The orthogonality of these two-qudit states is determined by
\begin{align}
\Tr\left(\bigg[\mathcal{B}^{a_1}_{b_1} \otimes \mathcal{B}^{a_2}_{b_2}\bigg] \bigg[ \mathcal{B}^{a^\prime_1}_{b^\prime_1} \otimes \mathcal{B}^{a^\prime_2}_{b^\prime_2}\bigg]\right)=0 \iff \begin{array}{c}
a_1=a_1^\prime\ \text{and}\ b_1\neq b_1^\prime \\
\text{or} \\
a_2=a_2^\prime\ \text{and}\ b_2\neq b_2^\prime
\end{array}
\end{align}

As before, we can associate with each state $\mathcal{B}^{a_1}_{b_1} \otimes \mathcal{B}^{a_2}_{b_2}$ a vertex $v(\{a_1,b_1\},\{a_2,b_2\})$ and the resulting orthogonality graph, $\Gamma_{\textrm{sep}}$, has edges
\begin{align}
v(\{a_1,b_1\},\{a_2,b_2\})\sim v(\{a_1^\prime,b_1^\prime\},\{a^\prime_2,b^\prime_2\}) \iff \begin{array}{c}
a_1=a_1^\prime\ \text{and}\ b_1\neq b_1^\prime \\
\text{or} \\
a_2=a_2^\prime\ \text{and}\ b_2\neq b_2^\prime
\end{array}
\end{align}
Note that this is exactly the graph that one would obtain by taking the graph product [see Equation~\eqref{eqn:ORdefn}] of $\Gamma_{\textrm{single}}$ with itself \emph{i.e}.,
\begin{align}
\Gamma_{\textrm{sep}}=\Gamma_{\textrm{single}} \textsc{ OR } \Gamma_{\textrm{single}}
\end{align}

With our notation defined as above, we can again describe relevant subsets of vertices \emph{i.e}.,
\begin{align}
\text{Maximum independent set:} \qquad \text{e.g., }\quad &\{v(\{a_1,1\},\{a_2,1\})|\ a_1,a_2 \in \{1,2,\ldots,d+1\}\}\\
\text{Maximum clique:} \qquad \text{e.g., }\quad &\{v(\{1,b_1\},\{1,b_2\})|\ b_1,b_2 \in \{1,2,\ldots,d\}\} \label{eqn:sepclique}
\end{align}

This implies that $\alpha(\Gamma_{\textrm{sep}})=(d+1)^2$ whereas $\omega(\Gamma_{\textrm{sep}})=d^2$. Furthermore, we can cover all $[d(d+1)]^2$ vertices of $\Gamma_{\textrm{sep}}$ using $(d+1)^2$ distinct, non-overlapping maximum cliques---simply by letting $a_1$ and $a_2$ vary over all possible values in $\{1,2,\ldots,d+1\}$ (in our example clique of Equation~\eqref{eqn:sepclique} we have set $a_1=a_2=1$). This last fact tell us that the clique cover number (see Section~\ref{sec:Graphtheoreticalprerequisites}) is exactly $\overline{\chi}(\Gamma_{\textrm{sep}})=(d+1)^2$. The following chain of inequalities holds for any graph $\Gamma$,
\begin{align}
\alpha(\Gamma) \leq \vartheta(\Gamma )\leq \alpha^*(\Gamma )\leq \overline{\chi}(\Gamma)
\end{align}
Combining this with the preceding discussion, we see that the {orthogonality} graph for separable states has parameters
\begin{align}
\alpha(\Gamma_{\textrm{sep}}) = \vartheta(\Gamma_{\textrm{sep}} )=\alpha^*(\Gamma_{\textrm{sep}} )=\overline{\chi}(\Gamma_{\textrm{sep}})=(d+1)^2
\end{align}
which implies that the separable subset on their own are insufficient to manifest contextuality.

\subsubsection{. Entangled Stabilizer States}

%

Here we will argue, {using purely graph-theoretical techniques, that} the set of entangled two-qudit stabilizer states $\{\Pi\}_{\textrm{ent}}$ does not exhibit state-independent contextuality{, when $d$ is an odd prime}. We have only shown this to be true for primes up to $11$ because one part of our argument still relies on brute force computation. What follows is not necessarily the easiest way of proving our claim, but our recognition of the orthogonality graph $\Gamma_{\textrm{ent}}$ as a Cayley graph is novel, and may be of independent~interest.

A \Jam state, $\ket{J_U}$, [see Equation (\ref{eqn:JamIso})] for which $U \in \mathcal{C}_d$ must be a maximally entangled, bipartite stabilizer state. Since the set of bipartite entangled stabilizer (BES) states is isomorphic to the Clifford group, we can hope to find some group-theoretical structure underpinning the orthogonality relationship between these BES states. Two BES states are orthogonal if and only if their isomorphic Clifford gates are trace orthogonal
\begin{align}
\langle J_C \vert J_{C^\prime} \rangle =0 \iff \Tr(C^\dag C^\prime)=0
\end{align}
Note that, by definition, $C^\dag C^\prime$ must itself be an element of the Clifford group. Define $T$ as the subset of Clifford operators having trace zero,
\begin{align}
T:=\{C \in \mathcal{C}_d \vert \Tr(C)=0 \}
\end{align}

Two bipartite entangled states $\ket{J_C}$ and $\ket{J_{C^\prime}}$ are adjacent in $\Gamma_{\textrm{ent}}$ if and only if the product $C^\dag C^\prime \in T$. Crucially, this is exactly the structure given to us by a Cayley graph introduced in Section~\ref{sec:Graphtheoreticalprerequisites}. Note that similar Cayley graphs have also been used in the creation of MUBs \cite{WvDMH:2011}.

To exploit this relationship we should first examine the structure of the subset $T$. We begin by defining the Legendre Symbol, $\ell_d$, as

\begin{align*}
\ell_d(x) =
\begin{cases}
\;\;\,1\;\;\, \text{ if } &x \text{ is a quadratic residue} \pmod{d} \\
-1\;\, \text{ if } &x \text{ is a quadratic non-residue} \pmod{d}\\
\;\;\,0\;\;\, \text{ if } &x \equiv 0 \pmod{d}
\end{cases}
\end{align*}
where a quadratic residue is an integer in $\Zd$ that is of the form $x^2 \bmod d$ for some $ x \in \Zd$. Next we quote a result from Appleby \cite{Appleby:arxiv09} that relates the trace of the unitary operator $C_{(F\vert u)}$ to its constituents $F$ and $u$
\begin{align}
(\text{ Case 1: }\qquad\beta =0 \Rightarrow \alpha\neq 0\ )\hspace{1cm} \nonumber \\
\lvert \text{Tr}\left(C_{(F\vert u)}\right) \rvert = \begin{cases} |\ell_d(\alpha)|=1 \quad &(\text{Tr}(F)\neq 2) \\
|\ell_d(\gamma)| \sqrt{d} \delta_{u_1,0}\quad &(\text{Tr}(F)= 2, \gamma \neq 0)\\
d \delta_{u_1,0}\delta_{u_2,0}\quad &(\text{Tr}(F)= 2, \gamma =0)
\end{cases}\label{eqn:tracerelations1} \\
(\text{ Case 2: }\qquad\beta \neq 0 \ )\hspace{2cm} \nonumber\\
\lvert \text{Tr}\left(C_{(F\vert u)}\right) \rvert = \begin{cases} |\ell_d(\text{Tr}(F)- 2)|=1 \quad &(\text{Tr}(F)\neq 2)\\
|\ell_d(-\beta)| \sqrt{d} \delta_{u_2,\beta^{-1}(1-\alpha)u_1 }\quad &(\text{Tr}(F)= 2)
\end{cases}\label{eqn:tracerelations2}
\end{align}

By examining Equations \eqref{eqn:tracerelations1} and \eqref{eqn:tracerelations2}, with particular attention to the three relevant cases satisfying $\Tr(F)=2$, we find that there are exactly
\begin{align}
|T|=d(d-1)^2+(d^2-1)+[d(d-1)]^2=[d(d-1)+1](d^2-1)
\end{align}
such $C_{(F\vert u)}$ with $\Tr\left(C_{(F\vert u)}\right)=0$. The structure of the traceless subset of Clifford gates is further simplified by using a result of Zhu \cite{Zhu:2010} that decomposes the Clifford group into conjugacy classes. Using the notation that the class $\left[C_{(F\vert u)}\right]$ with a representative element $C_{(F\vert u)}$ has size $\left|\left[C_{(F\vert u)}\right]\right|$, we have
\begin{align}
&\Bigg|\bigg[ C_{\left(\bigl[\begin{smallmatrix} 1 & 0\\ 0 & 1 \end{smallmatrix}\bigr]\middle\vert \bigl[\begin{smallmatrix} 1 \\0 \end{smallmatrix}\bigr]\right)} \bigg]\Bigg|=d^2-1 \\
&\Bigg|\bigg[ C_{\left(\bigl[\begin{smallmatrix} 1 & 0\\ 1 & 1 \end{smallmatrix}\bigr]\middle\vert \bigl[\begin{smallmatrix} u_1 \\0 \end{smallmatrix}\bigr]\right)} \bigg]\Bigg|=d(d^2-1) \qquad \forall\ u_1 \in \{1,2,\ldots,(d-1)/2\}\\
&\Bigg|\bigg[ C_{\left(\bigl[\begin{smallmatrix} 1 & 0\\ \nu & 1 \end{smallmatrix}\bigr]\middle\vert \bigl[\begin{smallmatrix} u_1 \\0 \end{smallmatrix}\bigr]\right)} \bigg]\Bigg|=d(d^2-1) \qquad \forall\ u_1 \in \{1,2,\ldots,(d-1)/2\}
\end{align}
where $\nu$ is a quadratic non-residue \emph{i.e}., $\nexists\ x \in \Zd$ such that $\nu=x^2 \bmod d$. By using these conjugacy classes to enumerate the number of traceless Clifford gates we find
\begin{align*}
|T|=(d^2-1)+2\left(\frac{d-1}{2}\right)d(d^2-1)=[d(d-1)+1](d^2-1)
\end{align*}
which confirms our previous counting argument.

The connection set $T$ is closed under conjugation so that $\Gamma(G,T)$ is a \emph{normal} Cayley graph. This implies \cite{Godsil:2008} that
\begin{align}
\text{ if }\ \alpha(\Gamma)\omega(\Gamma)=|\Gamma|\ \text{ then }\ \chi(\Gamma)=\omega(\Gamma)
\end{align}
Clearly, $\{\Pi\}_{\textrm{ent}}$ contains a complete orthonormal basis of $d^2$ vectors, which implies that $\omega(\Gamma_{\textrm{ent}})=d^2$. In our case, the implication of finding $\alpha(\Gamma)=p(p-1)$ is that $\chi(\Gamma)=D=d^2$ and therefore the set $\{\Pi\}_{\textrm{ent}}$ does not exhibit state-independent contextuality. As is well known, the problem of finding a maximum independent set or the chromatic number of a graph is NP-hard in both cases. However, currently available implementations for clique-finding, e.g., Cliquer \cite{Ostergard:2002, Ostergard:2002-1}, seem to perform much better than algorithms for vertex coloring, e.g., Mathematica \cite{MMA}. For all odd primes $p\in \{3,5,7,11\}$ that we have checked we can find \cite{Ostergard:2002, Ostergard:2002-1} independent sets of size
\begin{align}
\alpha(\Gamma_{\textrm{ent}})=p(p^2-1)
\end{align}
so that the set $\{\Pi_{\textrm{ent}}\}$ does not satisfy the $\chi(\Gamma)>D$ criterion \cite{Cabello:arxiv2011} for state-independent contextuality. Only for the qubit case does $\{\Pi_{\textrm{ent}}\}$ exhibit state independent contextuality \emph{i.e}.,
\begin{align}
d=2: \qquad \alpha(\Gamma_{\textrm{ent}})=5, \quad \chi(\Gamma_{\textrm{ent}})=5
\end{align}
{Given these results, it seems natural to} conjecture that $d=2$ is the only prime qudit dimension for which the entangled set $\{\Pi_{\textrm{ent}}\}$ displays state-independent contextuality. It would be nice to find a {purely graph-theoretical} proof that state-independent contextuality is impossible for all odd prime dimensions, \emph{i.e}., without resorting to numerics.

Associated with each row or column of the Peres--Mermin square in Table \ref{tab:PMsquare} is a complete orthonormal basis of four entangled stabilizer states. For example, the basis associated with Row 1 can be decomposed as the following four projectors
\begin{align}
\text{Row 1:} \begin{cases} \frac{1}{4}(I\otimes I+X\otimes Y+Y\otimes X+Z\otimes Z)=\Pi_1\\
\frac{1}{4}(I\otimes I-X\otimes Y+Y\otimes X-Z\otimes Z)=\Pi_2\\
\frac{1}{4}(I\otimes I+X\otimes Y-Y\otimes X-Z\otimes Z)=\Pi_3\\
\frac{1}{4}(I\otimes I-X\otimes Y-Y\otimes X+Z\otimes Z)=\Pi_4 \end{cases}
\end{align}
and similarly for all other rows and columns, eventually producing a set of $24$ projectors $\{\Pi_1,\ldots, \Pi_{24}\}$, which provide a proof of the Kochen--Specker theorem. That the Peres--Mermin square can be recast as a statement about $24$ projectors is well known \cite{Peres:1991}. It seems to have gone unnoticed, however, that the particular choice of Peres--Mermin magic square in Table \ref{tab:PMsquare} is completely equivalent to $\Gamma_{\textrm{ent}}$ in the way we have just described.


\subsubsection{. Generic Impossibility of Qudit State-Independent Contextuality}\label{sec:Combined separable and entangled stabilizer states}

With reference to Tables \ref{tab:GraphAlphas} and \ref{tab:GraphExpectations} we see that the condition for contextuality, $\alpha(\Gamma_{\textrm{tot}})<\langle \Sigma_{\Gamma_{\textrm{tot}}} \rangle^{\textsc{QM}}$, holds for the two-qubit case, but not for the two-qudit case when $d=3$ or $d=5$. The proofs are purely computational so we omit any further details. The results obtained {so far} prompt the question whether state-independent contextuality is ever possible using odd prime dimensional qudits. {In fact this question has already been answered in previous works that dealt with a discrete quasi-probabilistic representation of odd-dimensional quantum operators. Implicit in the discrete Wigner function (DWF) construction of Gross \cite{Gross:2006} is a local hidden variable model of a sub-theory of quantum mechanics, which was made explicit in a more recent work \cite{Veitch:2012}. This sub-theory is defined by restricting the available states and effects (\emph{i.e}., operations corresponding to either measurements or transformations) to those that are positively represented within the DWF. Most relevant for our discussion is the fact that this sub-theory includes all stabilizer measurements and all (mixtures of) stabilizer states. In particular, a set of odd-dimensional stabilizer measurements applied to the completely mixed state always has a local hidden variable model. Viewed another way, this tells us that state-independent contextuality is impossible whenever qudit (with odd $d$) stabilizer measurements are used. State-dependent contextuality using odd $d$-dimensional stabilizer projectors is still possible, however, so long as the measured state is not positively represented. This is the case in the examples presented in the next section. }

\subsection{State-Dependent Contextuality Using Bell Inequalities}\label{sec:State-dependent contextuality using Bell inequalities}

In the previous section we sought cases for which a sum of projectors $\Sigma_{\Gamma}$ had an expectation value $\Tr(\rho\Sigma_{\Gamma}) > \alpha(\Gamma)$, regardless of the state $\rho$ that was used. In this section we will construct non-contextuality inequalities that are violated by only some quantum states, \emph{i.e}., only some states are contextual with respect to this inequality. We will convert qudit Bell inequalities into {orthogonality} graphs, creating non-contextuality inequalities that can only be violated by entangled states. Starting from the famous two-qubit CHSH inequality \cite{CHSH:1969} e.g.,
\begin{align}
\langle \mathcal{B} \rangle\leq 2\quad \text{ where }\quad \mathcal{B}=X\otimes X+X\otimes Y+Y\otimes X-Y\otimes Y \label{eqn:origCHSH}
\end{align}
we will show how this corresponds to a non-contextuality inequality in the sense outlined in Section~\ref{sec:Quantum Contextuality and Graph Theory}. {A similar analysis has already been done numerous times in the literature (see e.g., \cite{CSW:arxiv2010,Sadiq:2013})}. Subsequently, we describe the resulting graphs and non-contextuality inequalities that arise when a qudit (for odd prime $d$) version of the CHSH inequality is used. {This is our novel contribution.}

In order to see the graph theoretic interpretation of Equation~\eqref{eqn:origCHSH}, first rewrite $\mathcal{B}$ as follows;
\begin{align}
\mathcal{B} &= X\otimes X+X\otimes Y+Y\otimes X-Y\otimes Y \label{eqn:CHSH_decomp} \\
&=2\left(\Pi_{(1,1|,0,0)[0]} +\Pi_{(1,1|,0,1)[0]} +\Pi_{(1,1|,1,0)[0]} +\Pi_{(1,1|,1,1)[1]} -2\mathbb{I}_4 \right)\\
&=2\big( \sum_{a+b=0}\left(\Pi_{(1|0)[a]}\otimes\Pi_{(1|0)[b]}\right) +\sum_{a+b=0}\left(\Pi_{(1|0)[a]}\otimes\Pi_{(1|1)[b]}\right)\nonumber \\ &\quad +\sum_{a+b=0}\left(\Pi_{(1|1)[a]}\otimes\Pi_{(1|0)[b]}\right) +\sum_{a+b=1}\left(\Pi_{(1|1)[a]}\otimes\Pi_{(1|1)[b]}\right) -2\mathbb{I}_4 \big) \label{eqn:sumof8}
\end{align}
where the second equality is a consequence of the definition Equation~\eqref{eqn:TwoQProj}, while the final equality is a consequence of Equation~\eqref{eqn:SumRank1}. Combining $\langle \mathcal{B} \rangle\leq 2$ with Equation~\eqref{eqn:sumof8} and denoting the sum of eight projectors as $\Sigma_{\Gamma_{\textrm{CHSH}}}$ returns the following inequality
\begin{align}
\langle \mathcal{B} \rangle&\leq 2\\
\Rightarrow \langle 2\Sigma_{\Gamma_{\textrm{CHSH}}}-4\mathbb{I}_4 \rangle&\leq 2 \\
\Rightarrow \langle \Sigma_{\Gamma_{\textrm{CHSH}}} \rangle&\leq 3 \label{eqn:CHSHasNCI}
\end{align}
where the latter is of the correct form for a non-contextuality inequality.

\begin{figure}[H]
\begin{center}
\includegraphics[scale=0.5]{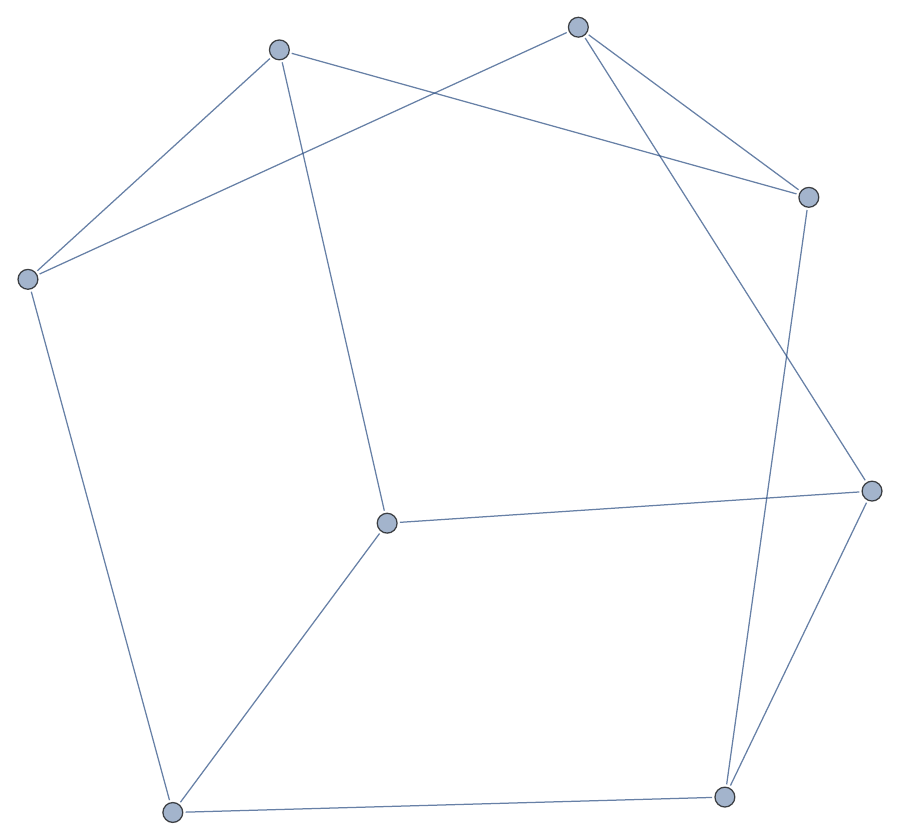}
\end{center}\caption{\label{fig:Mobius4} The graph $\Gamma_{\textrm{CHSH}}$ that arises from the orthogonality relationship between the set of projectors $\{\Pi_1,\ldots,\Pi_8\}$ as constructed in Equation~\eqref{eqn:sumof8}. The corresponding \mbox{non-contextuality} inequality $\langle \Sigma_\Gamma \rangle \leq 3$ is completely equivalent to the Bell inequality $\langle \mathcal{B} \rangle\leq 2$. }
\end{figure}

Each of the resulting eight projector terms in Equation~\eqref{eqn:sumof8} is a stabilizer projector of rank one \cite{GammaCHSH8}. By checking the commutation relations between projectors $\{\Pi_1,\ldots,\Pi_8\}$ we get a {orthogonality} graph ${\Gamma_{\textrm{CHSH}}}$ as depicted in Figure~\ref{fig:Mobius4}. The relevant graph-theoretic quantities are $\alpha({\Gamma_{\textrm{CHSH}}})=3, \vartheta({\Gamma_{\textrm{CHSH}}})=2+\sqrt{2}$ and $\alpha^*({\Gamma_{\textrm{CHSH}}})=4$. The first of these quantities tells us that the relevant non-contextuality inequality is
\begin{align}
\langle \Sigma_{\Gamma_{\textrm{CHSH}}} \rangle_{\max}^{\textsc{NCHV}}&\leq 3 \label{eqn:CHSH_max_leq_3}
\end{align}
which is exactly what our rewriting of the Bell inequality in Equation~\eqref{eqn:CHSHasNCI} told us.

In order to generalize the above example to qudits, we first rewrite the $d=2$ CHSH inequality in the following form, where as usual $\omega=e^{\tfrac{2 \pi i}{d}}$,
\begin{align}
\langle \mathcal{B}\rangle \leq 2\ \text{ where }\ \mathcal{B}=\sum_{j,k \in \mathbb{Z}_2}\omega^{jk}A_j\otimes B_k \quad\left(A_0,B_0=X, A_1,B_1=Y\right) \label{eqn:BellQubit}
\end{align}
Ji \emph{et al.} \cite{Ji:2008} (with a subsequent reexamination by Liang \emph{et al.} \cite{Liang:2010}) have shown that the form of the Bell inequality in Equation~\eqref{eqn:BellQubit} can be generalized to work for qudits too. They introduced a natural modification of the sum comprising the Bell operator, along with a careful choice of measurement operators e.g.,
\begin{align}
A_j=\omega^{j(j+1)}XZ^j,\ B_k=\omega^{(2^{-1})^2(k^2+2k)}XZ^{2^{-1}k}
\end{align}

For $d=3$ \emph{i.e}., qutrits we have
\begin{align}
\langle \mathcal{B}\rangle \leq 9\ \text{ where }\ \mathcal{B}=\sum_{\substack{n \in \mathbb{Z}^*_3\\ j,k \in \mathbb{Z}_3}}\omega^{njk}A^n_j\otimes B^n_k \label{eqn:QutritCHSH}
\end{align}
With this information we can rewrite
\begin{align*}
\mathcal{B}=3\big(
\Pi_{(1,1|0,0)[0]}+\Pi_{(1,1|0,1)[0]}+\Pi_{(1,1|0,2)[1]}+
\Pi_{(1,1|1,0)[0]}+\Pi_{(1,1|1,1)[1]}\\+\Pi_{(1,1|1,2)[0]}
+\Pi_{(1,1|2,0)[1]}+\Pi_{(1,1|2,1)[0]}+\Pi_{(1,1|2,2)[0]}-3\mathbb{I}_9\big) 
\end{align*}
Breaking up this expression even further into $27$ distinct rank-one projectors, via Equation~\eqref{eqn:SumRank1}, we get a 10-regular graph $\Gamma_{\textrm{CHSH}}$ of order $27$ \cite{GammaCHSH27}, with $\alpha(\Gamma_{\textrm{CHSH}}) =6$ and $\vartheta(\Gamma_{\textrm{CHSH}}) \approx 7.098 $. The corresponding non-contextuality inequality is thus
\begin{align}
\langle \Sigma_{\Gamma_{\textrm{CHSH}}} \rangle_{\max}^{\textsc{NCHV}}&\leq 6
\end{align}
which is easily seen to be completely equivalent to the original inequality $\langle \mathcal{B}\rangle \leq 9$ in Equation~\eqref{eqn:QutritCHSH}. By looking at the spectrum of the operator $\Sigma_{\Gamma_{\textrm{CHSH}}}$, we see that the maximum achievable using these projectors is $\lambda _{\max }(\Sigma_{\Gamma_{\textrm{CHSH}}})=6.4115$, which falls some way short of the $\langle \Sigma_{\Gamma_{\textrm{CHSH}}} \rangle_{\max}^{\textsc{QM}}=7.0981 $ that is achievable if we are allowed complete generality in our set of projectors $\{\Pi_1,\Pi_2,\ldots,\Pi_{27}\}$.

For $d=5$ we get a CHSH inequality of the form
\begin{align}
\langle \mathcal{B}\rangle \leq 35\ \text{ where }\ \mathcal{B}=\sum_{\substack{n \in \mathbb{Z}^*_5\\ j,k \in \mathbb{Z}_5}}\omega^{njk}A^n_j\otimes B^n_k 
\end{align}
where the Bell operator $\mathcal{B}$ eventually produces $125$ distinct rank-one projectors leading to a 36-regular graph ${\Gamma_{\textrm{CHSH}}}$ of order $125$, with parameters $\alpha=12, \vartheta \approx 18.09$ and $\lambda _{\max }\approx 13.09$. For $d=7$ we get an operator of the form
\begin{align}
\langle \mathcal{B}\rangle \leq 84\ \text{ where }\ \mathcal{B}=\sum_{\substack{n \in \mathbb{Z}^*_7\\ j,k \in \mathbb{Z}_7}}\omega^{njk}A^n_j\otimes B^n_k 
\end{align}
which eventually produces $343$ distinct rank-one projectors leading to a $78$-regular graph with parameters $\alpha=19$ and $\lambda _{\max }\approx 19.4112$. We were unable to calculate $\vartheta({\Gamma_{\textrm{CHSH}}})$ {for $d=7$} due to memory constraints. All these quantities are summarized in Table~\ref{tab:GraphCHSH}.

Observe that for all $d \in \{2,3,5,7\}$ that we have discussed, we see some structure in the graphs $\Gamma_{\textrm{CHSH}}$ obtained from qudit CHSH inequalities. Whereas it is quite easy to see that all graphs will have order $|\Gamma_{\textrm{CHSH}}|=d^3$, we conjecture based on the examples here that
\begin{align}
\Gamma_{\textrm{CHSH}} \text{ is a }(2d-1)(d-1)\text{{-regular graph}}
\end{align}

In the qubit case, the graph $\Gamma_{\textrm{CHSH}}$ is known to be a member of some well-studied families of graphs \emph{i.e}., $\Gamma_{\textrm{CHSH}}$ is a $(1,4)$-circulant graph and a $4$-M\"{o}bius ladder \cite{Cabello:2013}. It would be nice to see if $\Gamma_{\textrm{CHSH}}$ is expressible in a similar fashion for odd qudit dimensions.

\begin{table}[H]
\centering
\begin{tabular}{cccccc}
\toprule
& \boldmath$|\Gamma_{\textrm{CHSH}}|$ & \boldmath$\alpha(\Gamma_{\textrm{CHSH}})$ & \boldmath$\lambda_{\max}(\Sigma_{\Gamma_{\textrm{CHSH}}})$ & \boldmath$\vartheta(\Gamma_{\textrm{CHSH}})$ & \boldmath$C_{2k+1} \subset \Gamma_{\textrm{CHSH}}$ \\
\midrule $d=2$ & 8 & 3 & 3.414 & 3.414 & $k=2$\\
$d=3$ & 27 & 6 & 6.412 & 7.098 &$2 \leq k \leq 4$ \\
$d=5$ & 125 & 12 & 13.090 & 18.090 &$2 \leq k \leq 6$ \\
$d=7$ & 343 & 19 & 19.411 & ? &$2 \leq k \leq 10$\\
$d$ & $d^3$ & ? & ? & ? & ? \\
\bottomrule
\end{tabular}
\caption{\label{tab:GraphCHSH} Properties of the orthogonality graphs $\Gamma_{\textrm{CHSH}}$ that we have constructed by decomposing a two-qudit Bell inequality into stabilizer projectors. Applying $\langle \Sigma_{\Gamma} \rangle^{\textsc{NCHV}}_{\max}=\alpha(\Gamma)$ to these graphs gives the maximum achievable value of $\langle \Sigma_{\Gamma} \rangle$ over all local hidden variable theories. If we are allowed complete freedom in our choice of projectors then the maximum value of $\langle \Sigma_{\Gamma} \rangle$ achievable in quantum mechanics is given by $\vartheta(\Gamma)$. The maximum achievable value of $\langle \Sigma_{\Gamma_{\textrm{CHSH}}} \rangle$ using stabilizer projectors is denoted by $\lambda_{\max}(\Sigma_{\Gamma_{\textrm{CHSH}}})$. We do not know of any general solution for these parameters, nor were we able to calculate the Lovasz theta number of $\Gamma_{\textrm{CHSH}}$ for $d=7$, and so these quantities are denoted by question marks. The final column concerns subgraphs that can be found within $\Gamma_{\textrm{CHSH}}$, as discussed in Section~\ref{sec:Induced Subgraphs and equivalence of non-isomorphic graphs}.
}
\end{table}

%

\subsection{Induced Subgraphs and Equivalence of Non-Isomorphic Graphs}\label{sec:Induced Subgraphs and equivalence of non-isomorphic graphs}

In this section we discuss two additional features related to orthogonality graphs that \linebreak display~contextuality.

\begin{figure}[H]
\centering
\includegraphics[scale=1]{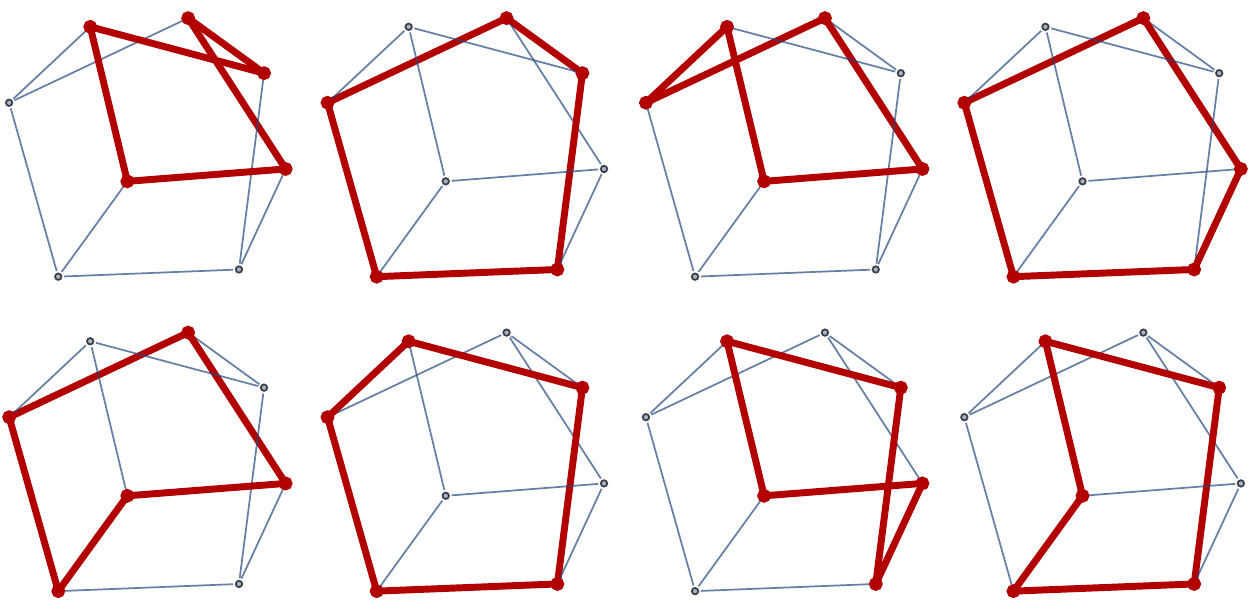}
\caption{\label{fig:G8allG5sub} All graphs that exhibit contextuality contain odd cycles $C_{2k+1}$ or their complements $\overline{ C_{2k+1}}$ \cite{Cabello:arxiv12}. For the case of the qubit CHSH graph, the only such subgraph that can be found is the pentagon $C_5$. There are $8$ distinct induced pentagons within the qubit CHSH graph, as depicted. }
\end{figure}

Firstly, we note a result due to Cabello \emph{et al.} \cite{Cabello:arxiv12}, which says that any graph displaying quantum contextuality must contain (as an induced subgraph) an odd cycle $C_{2k+1}$ or its complement $\overline{ C_{2k+1}}$ for some integer value of $k>1$. This is a necessary, but not sufficient, condition for a given graph $\Gamma$ to exhibit contextuality. In Table II of \cite{Cabello:arxiv12} the induced $C_{2k+1}$ or $\overline{ C_{2k+1}}$ associated with well-known Kochen--Specker sets are tabulated. Since the generalized CHSH graphs that we have discussed are new, we have performed a similar calculation and summarized the results in the final column of Table~\ref{tab:GraphCHSH}. All relevant subgraphs associated with the qubit CHSH inequality (and its corresponding graph $\Gamma_{\textrm{CHSH}}$) are depicted in Figure~\ref{fig:G8allG5sub}.

The second aspect of graph-based contextuality that we would like to note is the possibility of degeneracy or non-uniqueness in orthogonality graphs. We provide an example of two non-isomorphic orthogonality graphs that are both equivalent to the same Bell inequality. This non-unique association between inequalities and orthogonality graphs is a general feature, {and has already been noted elsewhere in the literature, e.g., \cite{Sadiq:2013}, but we find it instructive to provide an explicit example using the qubit CHSH~inequality.}

\begin{figure}[H]
\centering
\subfigure[]{\includegraphics[bb=280 280 550 500, clip=true,scale=0.95]{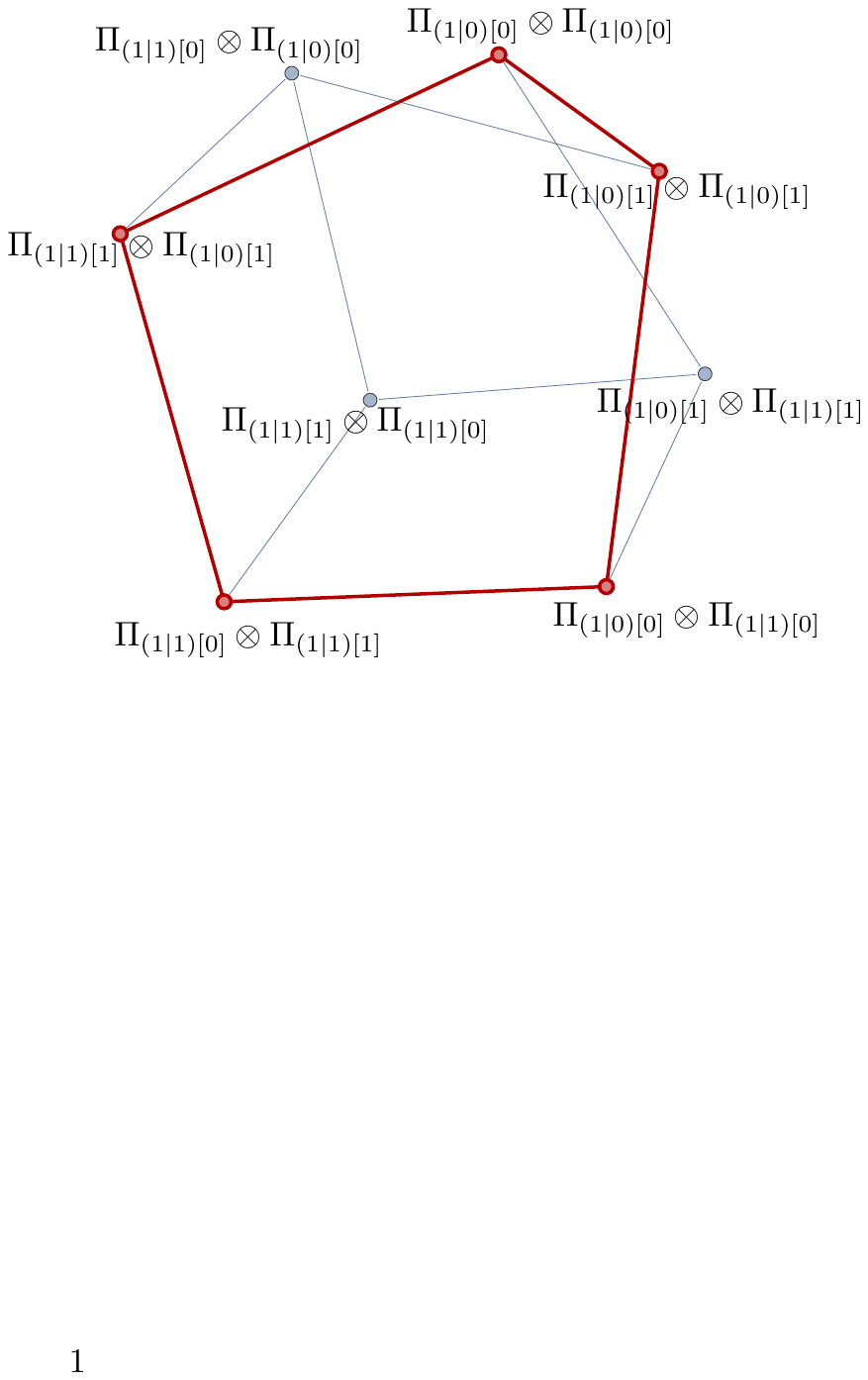}}\subfigure[]{\includegraphics[bb=290 280 550 500, clip=true,scale=0.95]{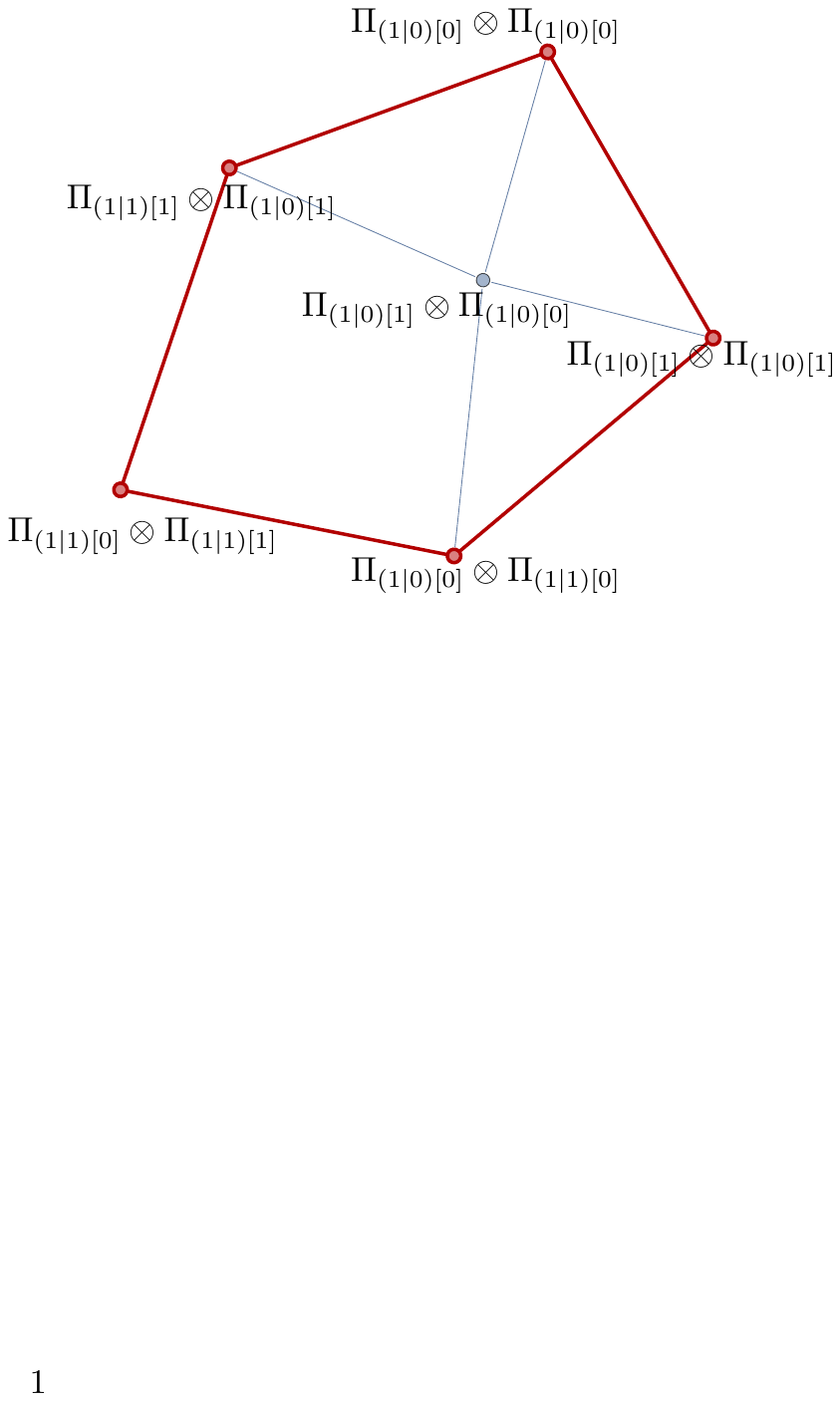}}
\caption{\label{fig:LabelledG8G6} Two non-isomorphic orthogonality graphs, with vertices explicitly labelled, that are both equivalent to CHSH Bell inequality $\protect{\langle X\otimes X+X\otimes Y+Y\otimes X-Y\otimes Y\rangle \leq 2}$. We showed in Equations~\eqref{eqn:CHSH_decomp}--\eqref{eqn:CHSH_max_leq_3} how application of the CSW formalism \cite{CSW:arxiv2010} to Figure~\ref{fig:LabelledG8G6}a demonstrated this equivalence. A similar argument for Figure~\ref{fig:LabelledG8G6}b is provided in Equations~\eqref{eqn:CHSH_decomp_again}--\eqref{eqn:CHSH_max_leq_2} below. Since our claim is that these graphs display contextuality, they must include an odd cycle or its complement; hence we highlight the pentagonal graph $C_5$ contained within both graphs. (\textbf{a}) The graph $\Gamma_{\textrm{CHSH}}$ comprised of $8$ projectors; (\textbf{b}) The graph $\Gamma^\prime_{\textrm{CHSH}}$ comprised of $6$ projectors.}
\end{figure}

\begin{figure}[H]
\centering
\includegraphics[scale=1.1]{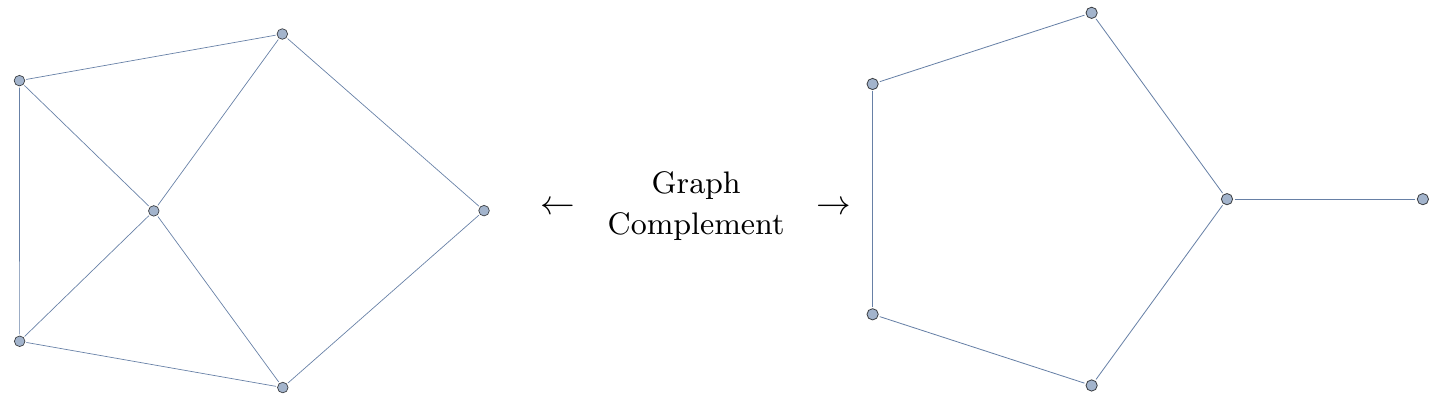}
\caption{\label{fig:Combined5Pan} The structure of the alternate CHSH contextuality graph $\Gamma^\prime_{\textrm{CHSH}}$ of Figure \ref{fig:LabelledG8G6}b can be recognized as the complement of the $5$-pan graph. The family of $n$-pan graphs consist of an $n$-cycle in addition to a single vertex that is connected to the $n$-cycle by a single edge.}
\end{figure}

By a direct calculation, one can show that the sum of six projectors given in Figure~\ref{fig:LabelledG8G6}b produces an operator $\Sigma_{\Gamma^\prime_{\textrm{CHSH}}}=(X\otimes X+X\otimes Y+Y\otimes X-Y\otimes Y+6\mathbb{I}_4)/4$, so that rewriting the CHSH inequality leads to
\begin{align}
\langle \mathcal{B} \rangle&\leq 2 \label{eqn:CHSH_decomp_again}\\
\Rightarrow \langle 4\Sigma_{\Gamma^\prime_{\textrm{CHSH}}}-6\mathbb{I}_4 \rangle&\leq 2 \\
\Rightarrow \langle \Sigma_{\Gamma^\prime_{\textrm{CHSH}}} \rangle&\leq 2 \label{eqn:CHSHasNCIagain}
\end{align}
where the latter is of the correct form for a non-contextuality inequality. The relevant graph-theoretic quantities are $\alpha({\Gamma^\prime_{\textrm{CHSH}}})=2, \vartheta({\Gamma^\prime_{\textrm{CHSH}}})=\sqrt{5}\approx 2.236$ and $\alpha^*({\Gamma^\prime_{\textrm{CHSH}}})=5/2$. The first of these quantities tells us that the relevant non-contextuality inequality is
\begin{align}
\langle \Sigma_{\Gamma^\prime_{\textrm{CHSH}}} \rangle_{\max}^{\textsc{NCHV}}&\leq 2 \label{eqn:CHSH_max_leq_2}
\end{align}
which confirms the equivalence with the original Bell CHSH inequality. In contrast to the original graph $\Gamma_{\textrm{CHSH}}$, the maximal quantum value of $\vartheta({\Gamma^\prime_{\textrm{CHSH}}})$ can never be reached using separable projectors, as this would violate Tsirelson's bound. The structure of this alternate graphical representation of the CHSH inequality can be understood as the graph complement of so-called $5$-pan graph depicted in Figure~\ref{fig:Combined5Pan}. {We have not seen this explicit description elsewhere in the literature but it is implied by the techniques used in \cite{Sadiq:2013}}.

Perhaps this non-uniqueness can be used to simplify, or better understand the structure of, the graphical representation of qudit Bell inequalities that we discussed in Section~\ref{sec:State-dependent contextuality using Bell inequalities}.



\section{Summary}

We have applied the graph-based contextuality formalism of Cabello, Severini and Winter \cite{CSW:arxiv2010} to sets of two-qudit stabilizer states. Sets of states that arise naturally because of structural or \linebreak group-theoretical significance are seen to display state-independent contextuality only if the particles are qubits, as opposed to qudits with prime dimension $d>2$. {While our use of graph-theoretical techniques was insufficient to prove that this should always hold for particles of prime dimension $d$, we noted a more general result based on the discrete Wigner function that implied that this is indeed the case. } It was already known that the two-qubit Bell CHSH inequality could be recast as a non-contextuality inequality with a related orthogonality graph \cite{CSW:arxiv2010}. Here we have transformed a qudit analogue of Bell CHSH inequality into an orthogonality graph and derived the related non-contextuality inequality and graph parameters. Along the way we have highlighted various structural properties of the orthogonality graphs under consideration.


\section*{Acknowledgements}
\vspace{12pt}

M.H. was financially supported by the Irish Research Council (IRC) as part of the Empower Fellowship program. J.V. acknowledges support from Science Foundation Ireland under the Principal Investigator Award No. 10/IN.1/I3013. {We thank Joseph Emerson for helpful comments on a previous version of this manuscript.}

\section*{Conflict of Interest}
\vspace{12pt}

The authors declare no conflict of interest.

\bibliographystyle{mdpi}
\makeatletter
\renewcommand\@biblabel[1]{#1. }
\makeatother

\begin{thebibliography}{----}
\bibitem{GammaCHSH8}
The 8 distinct projectors, in pairs associated with each \mbox{summation} in Equation~(\ref{eqn:sumof8}), are equivalent to following un-normalized vectors $ \{[(1,1,1,1),(1,-1,-1,1)],[(1,i,1,i),(1,-i,-1,i)],[(1,1,i,i), (1,-1,-i,i)],$\linebreak $ [(1,-i,i,1),(1,i,-i,1)]\}.$
\end{thebibliography}

\end{document}